\documentclass[prb,twocolumn,showpacs]{revtex4}
\usepackage{graphicx}
\usepackage{times}

\newcommand{\dist}{d}
\newcommand{\thick}{h}
\newcommand{\width}{w}
\newcommand{\micron}{\mu{\rm m}}

\usepackage{bm}

\DeclareBoldMathCommand{\bfmu}{\mu}

\begin{document}

    \title{Magnetic noise around metallic microstructures}
    \author{Bo Zhang and C. Henkel%
    \email{carsten.henkel@uni-potsdam.de}
    }
    \affiliation{
Institut f\"{u}r Physik, Universit\"{a}t Potsdam, 
    14469 Potsdam, Germany
}
    \date{28 Aug 2007}


\begin{abstract}
    We compute the local spectrum of the magnetic field near 
    a metallic microstructure at finite temperature. 
    Our main focus is on deviations from a plane-layered geometry 
    for 
    which we review the main properties. Arbitrary geometries are 
    handled with the help of numerical calculations based on surface 
    integral equations. The magnetic noise shows a significant 
    polarization anisotropy 
    above flat wires with finite lateral width, in stark contrast to 
    an infinitely wide wire. Within the limits of a two-dimensional 
    setting, our results provide accurate estimates for loss and 
    dephasing rates in so-called `atom chip traps' based on metallic
    wires.  A simple approximation based on the incoherent summation 
    of local current elements gives qualitative agreement with the 
    numerics, but fails to describe current correlations among 
    neighboring objects. 
\end{abstract}


\maketitle

\clearpage

\section{Introduction}


Thermal motion of charge carriers in a metallic object creates
a randomly fluctuating magnetic field in the object's vicinity.
These fields are relevant for many applications like high-precision
measurements of biomagnetic signals~\cite{Nenonen96}, nuclear magnetic 
resonance microscopy~\cite{Sidles00}, and miniaturized traps for
ultra cold atoms~\cite{Folman02,Fortagh07}.  
The original purpose of Purcell's influential
1946 paper~\cite{Purcell46} was to point out that these
magnetic fields have a spectral density that by far exceeds the Planck
law for blackbody radiation at low frequencies. In fact, if only
blackbody fields were present, magnetic dipole transitions between 
atomic or nuclear levels would never happen on laboratory time scales.  
The near fields sustained by material objects (that play the roles of
sources and cavity) give the dominant contribution.
Indeed, these near fields contain non-propagating (evanescent)
components that are thermally excited as well and that dominate over free 
space radiation~\cite{Agarwal75a,Varpula84}.
Phrased in another way, the dipole transition rate is
enhanced because the elementary excitations in the metal provide
additional decay channels~\cite{Chance78,Barnes98}.  


We focus in this paper on accurate calculations of magnetic field noise
that are able to describe objects of arbitrary shape. Such objects 
occur, for example, in magnetic microtraps where complex networks of
metallic wires create electromagnetic potentials with typical scales 
in the micron range~\cite{Folman02,Fortagh07}.  The behaviour of the
field spectrum, as the metallic geometry is changed, is far from
intuitive.The spectral density increases with the
material volume for small structures, but this trend saturates as soon as the
typical scale gets larger than the penetration length (skin depth) 
of the fields in the material.
It has even been found that a thin metallic layer can produce less 
noise than a half-space, depending on the ratios 
between observation distance, layer thickness, and skin 
depth~\cite{Varpula84,Scheel05a,Henkel05c}. 
Experiments in the field of biomagnetism have shown significant 
changes when a metallic film is cut, at constant volume, 
into stripes~\cite{Nenonen96}.
Calculations for these cases necessarily require numerical methods to 
describe the propagation of magnetic fields both in vacuum and inside 
a metallic structure. This is the main topic of this paper. We also 
discuss previously developed approximations for planar structures 
and within the magnetostatic regime where analytical calculations are 
possible. Our numerical methods are restricted here to two spatial 
dimensions (2D) where efficient solutions of Maxwell equations can be 
found with the help of boundary integral 
equations~\cite{Nieto,Harrington,Rockstuhl03,Rogobete04a}.
Our approach can also be combined with any other numerical method for 
field computations, permitting to cover three-dimensional cases as well.

The results we find can be summarized as follows. A planar 
structure (infinite lateral size) creates equal magnetic noise for all
components of the magnetic field vector. This does not apply in three
dimensions, but is specific to the two-dimensional setting we 
focus on here.
Finite metallic objects show a strong anisotropy: the noise occurs
preferentially along `azimuthal' directions circling around the 
object. Increasing the amount of metallic material does not always
give larger noise, in particular when the geometrical size becomes 
comparable to the skin depth.  We find qualitative agreement with 
measurements of Ref.\onlinecite{Nenonen96} where thermal field 
fluctuations are reduced when a metallic object is split into 
disconnected pieces.  The surface impedance approximation, that
provides an accurate description of metallic reflectors for
far-field radiation, is shown to be not reliable for observation
distances shorter than the skin depth.  Finally, qualitative
(albeit not quantitative)
agreement is obtained between our numerical data and an approximation
based on the incoherent summation of fields generated by thermal 
current elements filling up the metallic volume. This method has
been used in the interpretation of previous 
experiments~\cite{Vuletic04,Dikovsky05}. Our results are, to our 
knowledge, the first quantitative test of this approximation in a 
nontrivial geometry.

The paper is organized as follows. We first review the link between
the thermal radiation spectrum and classical dipole radiation
(Sec.\ref{s:stat-phys}). Planar structures are analyzed in 
Sec.\ref{s:layer} using the angular spectrum representation. We 
demonstrate
in particular the isotropy of the magnetic noise spectrum and discuss 
the accuracy of the surface impedance approximation. 
Sec.\ref{s:finite-wire} is devoted to our numerical scheme and to the
results for single and multiple objects of rectangular shape.

%
%

\section{Magnetic near field noise}
\label{s:stat-phys}

\subsection{Local noise power}

The fluctuations of the thermal 
magnetic field ${\bf B}( {\bf r}, t )$ 
are characterized by their local power spectral density (the Fourier 
transform of the autocorrelation function)
\begin{equation}
    {\cal B}_{ij}( {\bf r}; \omega ) = 
    \int\!{\rm d}\tau\,
    {\rm e}^{ {\rm i} \omega \tau }
    \langle {B}_{i}( {\bf r}, t )
    {B}_{j}( {\bf r}, t + \tau ) \rangle.
    \label{eq:def-mag-spectrum}
\end{equation}
Higher moments are not needed for our purposes, and the average
field (at frequency $\omega$) vanishes as is typical for thermal
radiation. We assume the 
field to be statistically stationary so that the 
spectrum~(\ref{eq:def-mag-spectrum}) does not depend on $t$. The 
diagonal tensor components 
${\cal B}_{ii}( {\bf r} ; \omega )$ 
give the spectrum for a given cartesian component $B_i( {\bf r} )$
or polarization direction. Previous 
work has shown a strong dependence on the position ${\bf r}$ near a 
metallic microstructure: power laws being the typical behaviour in 
the frequency range where the wavelength $\lambda = 2\pi c / \omega$ is 
much larger than the typical distances. For the temperature dependence,
see Eq.(\ref{eq:FD-theorem}) below. Our parameters of interest 
are: normal metallic conductors with temperatures above a few K,
${\bf r}$ in the micron range and $\lambda$ of the order of 
centimeters or larger ($\omega/2\pi \le 10\,{\rm GHz}$). This upper 
limit on frequency corresponds to the strong magnetic dipole transitions 
in typical alkali atoms.) In this regime, the frequency 
dependence of the noise spectrum is weak and occurs via the material 
response (permittivity $\varepsilon( \omega )$). A 
characteristic length scale is the field penetration length (skin
depth) $\delta$ defined by
\begin{equation}
    \frac{ 1 }{ \delta } = \frac{ 2 \pi }{ \lambdaÊ} {\rm Im} \sqrt{ 
    \varepsilon( \omega ) } = 
    \sqrt{ {\textstyle\frac12} \mu_{0} \omega \sigma( \omega ) }
\label{eq:def-skin-depth}
    %
\end{equation}
where $\sigma( \omega )$ is the conductivity and $\mu_{0}$ the vacuum 
permeability. The second expression is based on the Hagen-Rubens
approximation 
${\rm Im}\,\varepsilon( \omega ) \approx \sigma( \omega ) / (\varepsilon_{0} \omega )
\gg |{\rm Re}\,\varepsilon( \omega )|$. Within the Drude model for a 
metal, this is verified at frequencies much smaller than the 
charge carrier relaxation rate (in the $10^{15}\,{\rm 
s}^{-1}$ range at room temperature). For highly conducting 
materials (Au, Ag, Cu), this results in a skin depth of the order of
$100\,\micron (\omega / 2\pi\,{\rm MHz})^{-1/2}$. 
%
%
%
%
%

To compute the magnetic correlation spectrum, we use the 
fluctuation-dissipation theorem which is valid at thermal equilibrium
(temperature $T$)~\cite{Landau10}
\begin{equation}
    {\cal B}_{ij}( {\bf r}; \omega ) =
    \frac{ 2 \hbar }{
    {\rm e}^{ \hbar \omega / k_{B} T } - 1 }
    {\rm Im} \, {\cal G}_{ij}( {\bf r}, {\bf r}; \omega )
    \label{eq:FD-theorem}
\end{equation}
where ${\cal G}_{ij}( {\bf r}, {\bf r}'; \omega )$ is the Green
function for the magnetic field, i.e., the field generated at ${\bf
r}$ by a point magnetic dipole located at ${\bf r}'$ and oscillating
at the frequency $\omega$, $B_{i}( {\bf r}, t ) =
{\cal G}_{ij}( {\bf r}, {\bf r}'; \omega ) \mu_{j} \,{\rm e}^{ - {\rm
i} \omega t } + {\rm c.c.}$. This is actually a familiar result: the
imaginary part of ${\rm tr}\,{\cal G}( {\bf r}, {\bf r}; \omega )$ gives the
local density of magnetic field modes, and the temperature dependent
prefactor in Eq.(\ref{eq:FD-theorem}) their average occupation number.
The basic benefit of this formula is that it holds also for the full
correlation tensor and even near material objects that absorb the
field or generate thermal radiation. Generalizations to the
nonequilibrium case exist (fields produced by a `hot object'
surrounded by a `cold' environment)~\cite{Rytov3,Polder71},
but are not needed for our purposes (see remarks in Sec.\ref{s:conclusion}).
We also note that the temperature
dependent prefactor in Eq.(\ref{eq:FD-theorem}) reduces to
$2 k_{B} T / \omega$ for
$T \ge 0.1\,{\rm K}$.
In this
limit, the order of field operators in the correlation
function~(\ref{eq:def-mag-spectrum}) becomes irrelevant. (The order we
have adopted yields the rate of a magnetic dipole transition
$i \to f$ with
energy difference $E_{f} - E_{i} = \hbar \omega$.)

\subsection{Magnetic dipole radiation}

We are thus led to solve the following electrodynamic problem: find
the complex magnetic field amplitude ${\bf B}( {\bf r}; \omega | \bfmu )$
created by a monochromatic point dipole $\bfmu(t) = \bfmu\,{\rm e}^{
-{\rm i}\omega t} + {\rm c.c.}$ located at position ${\bf
r}'$. We then compute
\begin{equation}
G_{ij}( {\bf r}, {\bf r}'; \omega ) = \frac{ 
\partial B_i(  {\bf r}; \omega | \bfmu ) }{
\partial \mu_j }
\label{eq:def-Green-function-0}
\end{equation}
In the limit ${\bf r} \to {\bf r}'$ this field becomes the singular `self field'
and requires a cutoff in
wavevector space. Its imaginary part is cutoff-independent, however,
and given by
${\rm Im}\,{\bf B}( {\bf r}'; \omega ) = \mu_{0}\omega^3 \bfmu
/ (6\pi c^3)$ (in three-dimensional free space).

The field ${\bf B} = {\bf B}( {\bf r}; \omega | \bfmu )$ can be found
from the vector potential ${\bf A}$
that solves the inhomogeneous Maxwell equation
\begin{equation}
    \nabla \times \nabla \times {\bf A}
    - k_{0}^2 \varepsilon( {\bf r})
    {\bf A} = \mu_{0} \nabla \times \bfmu \delta( {\bf r} - {\bf r}' )
    ,
    \label{eq:Maxwell-H}
\end{equation}
where $k_{0} = \omega / c$.
The right-hand side is the current density 
corresponding to the magnetic dipole. There is no free charge 
density and we work in the gauge 
${\bf E} = {\rm i}\omega{\bf A}$.

\begin{figure}[hbt]
    \includegraphics*[width=80mm]{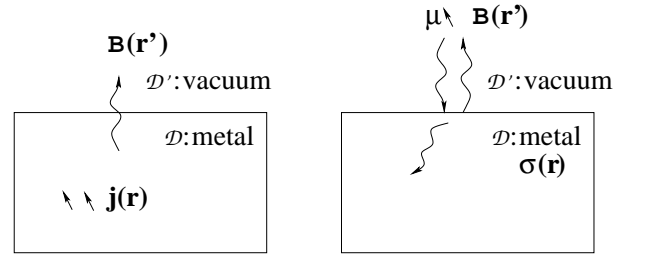}
    \caption[]{
    Sketch of the considered geometry: (left) current fluctuations
    in a microstructure generate magnetic field
    fluctuations ${\bf B}( {\bf r}' )$ at a position
    position ${\bf r}'$ outside it.
    (right) The magnetic noise spectrum is calculated
    from the magnetic field radiated by a point
    magnetic dipole $\bfmu$ located at ${\bf r}'$.
    ${\cal D}$ and ${\cal D}'$: domains where the
    conductivity $\sigma({\bf r}; \omega)$ is nonzero or zero,
    respectively.}
    \label{fig:sketch-wire}
\end{figure}

We now focus on the following geometry (fig.\ref{fig:sketch-wire}, right):
the position ${\bf r}'$ of the source (i.e., where the magnetic noise
spectrum is actually needed) is located in vacuum, and the metallic
microstructures are filling a domain ${\cal D}$ where ${\rm 
Im}\,\varepsilon( {\bf r}; \omega ) = \sigma( {\bf r}; \omega ) / 
(\varepsilon_0 \omega) $ is nonzero (and large).

The outside domain is called ${\cal D}'$. There, the vector 
potential satisfies an inhomogeneous Helmholtz equation
with wavenumber $k_{0}$.
All length scales we consider (distance dipole--microstructure 
$\dist$, object size) are much shorter than the 
wavelength so that $k_0$
is actually very small and can be neglected in a first 
approximation. This is the magnetostatic regime.
(The finite value of $k_{0}$ is, of course, at the origin 
of the nonzero magnetic LDOS 
in free space.)
We cannot make the magnetostatic approximation in 
${\cal D}$ because there, we have a
wavenumber $k_{0} \sqrt{\varepsilon( {\bf r}; \omega )} 
= (1+{\rm i}) / \delta( {\bf r} )$,
and the (local) skin depth $\delta( {\bf r} )$ is one of the 
characteristic length scales at hand.
%
%
%
%
%
%
%
%
%
%
The fields in the domains ${\cal D}$ and ${\cal D}'$
are connected by the usual matching conditions: the 
components of ${\bf A}$ tangential to the boundary are continuous,
and ${\bf B}$ is continuous (the material is 
non-magnetic). 

Eq.(\ref{eq:Maxwell-H}) provides
a unique solution subject to the boundary condition that at infinity, 
the field behaves like an outgoing wave. In three [two] dimensions, 
this corresponds 
to a vector potential proportional to 
${\rm e}^{{\rm i} k_0 s } / s$ [${\rm e}^{{\rm i} k_0 s } / \sqrt{ s }$]
in the free space domain ${\cal D}'$ when the distance $s = |{\bf r} - {\bf 
r}'| \to \infty$ to the source becomes large compared to $\lambda$. 
In the magnetostatic limit 
$k_0 \to 0$, the free space asymptotics is actually never reached 
at finite distances. The relevant boundary condition is then 
the same as for the scalar potential of an electric dipole: 
the vector potential goes to zero like $1/s^2$ [like $1/s$] in 
three [two] dimensions, respectively.

Since we deal with a metallic object with $|\varepsilon| \gg 1$, it is
tempting to perform the calculation based on the surface impedance
boundary condition.  The latter links the tangential components of
magnetic field and vector potential by $B_{t} = - {\rm i} \omega Z
A_{z}$,
where $\omega Z = 
(1+{\rm i})/\delta$.
Note that this is a local relation that can only hold if the scale of
variation of the fields on the object surface is much larger than
the skin depth $\delta$.  In the present study, a point-like source
illuminates the object with its near field [$A_{\rm bulk}(
{\bf r} - {\bf r}' )$ in Eq.(\ref{eq:bulk-solution})], and this
field shows a typical extension of the order of the object-source
distance $\dist$.  The surface impedance approximation is hence
expected to break down for $\dist \ll \delta$.  We shall confirm this
explicitly for the planar structures discussed in the following
Sec.\ref{s:layer}.

\section{Results: layer}
\label{s:layer}

\subsection{Two dimensions}

In this paper, we focus on a two-dimensional (2D) geometry to simplify the
numerical calculations described in Sec.\ref{s:finite-wire}.  
The magnetic moment is chosen in the computational plane (the
$xy$-plane), as shown in Fig.\ref{fig:sketch-wire}.  Adapting the wave
equation~(\ref{eq:Maxwell-H}) to two dimensions, we find that the
vector potential has a single nonzero component that points out of the
plane.  We then work with a scalar function $A( {\bf r} ) = A( x, y
)$ that solves
\begin{equation}
    \nabla^2 A +
    k_{0}^2 \varepsilon( {\bf r})  
    {A} = \mu_{0} \left( \mu_{y} \partial_{x'} - \mu_{x} \partial_{y'}
    \right) \delta( {\bf r} - {\bf r}' )
    \label{eq:Maxwell-A-2D}
\end{equation}
In a homogeneous medium (`bulk'), the solution with the appropriate
boundary conditions is
\begin{equation}
    A_{\rm bulk}( {\bf r} - {\bf r}'  ) = 
    \frac{ {\rm i} \mu_{0} }{ 4 } 
    \left( \mu_{y} \partial_{x'} - \mu_{x} \partial_{y'}
	\right)
	H_{0}( k_{0} \sqrt{ \varepsilonÊ} | {\bf r} - {\bf r}' | )
    \label{eq:bulk-solution}
\end{equation}
where $H_{0}$ is the Bessel function of the third kind (Hankel 
function), usually 
denoted $H_{0}^{(1)} = J_{0} + {\rm i} Y_{0}$. 
>>From this, we get the magnetic field by taking the `curl',
$B_{x} = \partial_{y} A$, $B_{y} = -\partial_{x} A$. 
The resulting self field in free space is
\begin{equation}
    {\rm Im}\,{\bf B}( {\bf r}' | \bfmu ) = {\textstyle\frac{ 1 }{ 8 }}
    \mu_{0} k_{0}^2 \,\bfmu
    \label{eq:free-space-self-field}
\end{equation}
provided the dipole $\bfmu$ is real. 
In the magnetostatic limit, this field is negligibly small.  The
bulk solution~(\ref{eq:bulk-solution}) then goes over into
\begin{equation}
    A_{\rm bulk}( {\bf r} - {\bf r}' ) \approx
	- \frac{ \mu_{0} }{ 2\pi } 
	\frac{ ( x - x' ) \mu_{y}  -  ( y - y' ) \mu_{x} }{
	| {\bf r} - {\bf r}' |^2 }
    \label{eq:mstat-bulk-solution}
\end{equation}
This equation describes the field with which the dipole `illuminates' 
the sample. Note that it is scale-free: the typical `spot size' on the 
microstructure is only determined by the distance $\dist$ between
dipole and top surface.

\subsection{Reflected field}

In this section, we consider that 
the boundary of the medium is the plane $y = 0$; the field at the source
point ${\bf r}' = (0,d)$ is then related to the (Fresnel) reflection coefficients
from the surface. We expand the solution to Eq.(\ref{eq:Maxwell-A-2D})
in plane waves (wavevector $k$ parallel to the
boundary) and have above the medium ($y > 0$):
\begin{eqnarray}
    A( x, y ) &=& 
    \mu _{0} ( \mu _{x} \partial _{y'} - \mu_{y}\partial_{x'} )
    \int\limits_{-\infty}^{+\infty}\!\frac{{\rm d}k}{ 2\pi }
    \frac{ {\rm e}^{{\rm i} k (x-x')} }{ 2 \kappa }
    \nonumber\\
    && \times
    \left(
    {\rm e}^{ -\kappa| y - y' | }
    +
    r( k ) {\rm e}^{ - \kappa( y + y' ) }
    \right)
    \label{eq:plane-wave-expansion}
\end{eqnarray}
where
$\kappa = \sqrt{ k^2 - k_{0}^2 }$.
(The square root is chosen such that ${\rm Re}\,\kappa \ge 0$.) 
The coefficient $r( k )$ describes the reflection of the field from 
the medium boundary. It is given by the Fresnel formula
\begin{equation}
    r(k) = 
    r_{\rm half\,space}( k ) \equiv
    \frac{ \kappa - \kappa_{\rm m} }{ \kappa + \kappa_{\rm m} }, \qquad
    \kappa_{\rm m} = \sqrt{ k^2 - 2{\rm i} / \delta^2 }
    \label{eq:half-space-r}
\end{equation}
for a medium with skin depth $\delta$ filling the half-space $y < 0$. For a 
layer (thickness $\thick$) on top of a substrate, we have
\begin{equation}
    r_{\rm layer}(k) = \frac{ r_{\rm top} + r_{\rm bottom} {\rm e}^{ -
    2\kappa_{\rm m} \thick } }{ 1 - r_{\rm top} r_{\rm bottom} {\rm e}^{ -
    2\kappa_{\rm m} \thick } }
    \label{eq:layer-r}
\end{equation}
where $r_{\rm top} = r_{\rm half\,space}$ is given by 
Eq.(\ref{eq:half-space-r}) and $r_{\rm bottom}$ 
describes the reflection from the layer--substrate interface. 
It is given by Eq.(\ref{eq:half-space-r}) with the replacements
$\kappa \mapsto \kappa_{\rm m}$, 
$\kappa_{\rm m} \mapsto \kappa_{\rm s} = 
(k^2 - \varepsilon_{\rm s} k_{0}^2 )^{1/2}$ where $\varepsilon_{\rm s}$ is 
the substrate permittivity.

All the relevant information for the magnetic noise power is contained
in the reflection coefficient $r( k )$.  In fact, when the integral in~(\ref{eq:plane-wave-expansion}) is performed and the imaginary part
taken, it turns out that the
reflected waves (second term) dominate over the free space contribution 
(first term) by at least
a factor $\lambda^2 \delta / \dist^3 \gg 1$.  This is connected to the
fact that the relevant wavenumbers $k$ for our problem are of the
order of $1/(y+y') = 1/(2\dist)$ which is much larger than $k_{0}$.  We can hence
apply the approximation $\kappa \approx |k|$.  The reflection
coefficient~(\ref{eq:half-space-r}) for the metallic half-space then
depends only on the parameter $k \delta$.  For the metallic layer
geometry, we focus for simplicity on a substrate whose conductivity is
much smaller than in the metal.  The influence of the substrate has
been studied in Ref.\onlinecite{Zhang05a}: already a ratio of 10 to
100 between the substrate and layer conductivities is sufficient to
make the substrate behave like vacuum.  We then have $r_{\rm bottom}
\approx - r_{\rm top}$ in Eq.(\ref{eq:layer-r}).

\subsection{Polarization dependence}
\label{s:isotropy}

Let us analyze first the dependence on the orientation of the source
dipole. If $\bfmu$ is perpendicular to the medium (only $\mu_{y} \ne 
0$), the reflected field is given by
\begin{equation}
B_{y}( {\bf r} | \mu_{y} ) 
    = \mu_{0} \mu_{y}
    \int\limits_{-\infty}^{\infty}\!\frac{{\rm d}k}{ 2\pi }
    \frac{k^2}{2\kappa}
    r( k ) 
    {\rm e}^{ {\rm i} k (x - x') }
    {\rm e}^{ - \kappa (y + y') }
    \label{eq:B-yy}
\end{equation}
The limit ${\bf r}  \to 
{\bf r}'$ yields an imaginary part
\begin{equation}
{\rm Im}\,B_{y}( {\bf r}' | \mu_{y} )
    =     \mu_{0} \mu_{y}  {\rm Im}\, 
    \int\!\frac{{\rm d}k}{ 2\pi }
   \frac{k^2}{2 \kappa} 
   r( k ) 
    {\rm e}^{ - 2 \kappa \dist }    .
    \label{eq:Im-B-yy}
\end{equation}
Repeating the calculation for a parallel dipole, we find for
${\rm Im}\,B_{x}( {\bf r}' | \mu_{x} )$ the same 
expression as Eq.(\ref{eq:Im-B-yy}), and consequently the
noise spectrum is isotropic, ${\cal B}_{xx} = {\cal B}_{yy}$. 
This is a remarkable property of a laterally infinite 
structure in 2D. (In 3D, the polarization perpendicular to 
a planar interface has a noise power twice as large as the
parallel polarization~\cite{Varpula84,Henkel99c}.)
We show below that a significant polarization
anisotropy arises above a metallic wire of finite width. 

\subsection{Wavevector dependence}

\begin{figure}[tbp]
    \centering
    \includegraphics*[width=80mm]{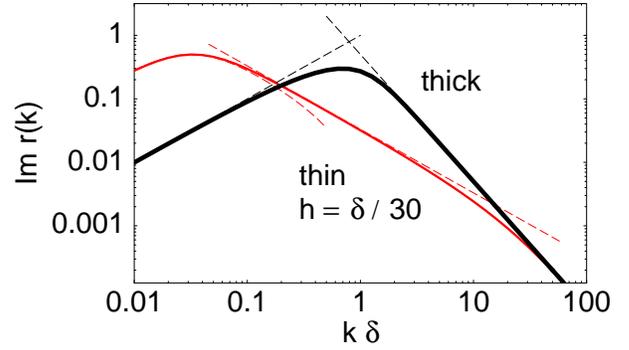}
    \caption[]{Reflection coefficients~(\ref{eq:half-space-r},
    \ref{eq:layer-r}) for thin and thick metallic layers.  We plot
    the imaginary part only.  The dashed lines represent the formulas of
    Table~\ref{t:asymptotics-r}.  The wavenumber is scaled to the
    inverse skin depth $1/\delta$. For the thick layer, $h =
    \infty$. We take the conductivity of gold at room temperature
    and a frequency $\omega/2\pi \approx 1.1\,{\rm MHz}$ in all
    figures. This leads to the value $\delta = 71 {\rm \mu m}$
    and a vacuum wavelength $\lambda / \delta \approx 3.8 \times 10^{6}$.}
    \label{fig:integrands}
\end{figure}

The reflection coefficient~(\ref{eq:layer-r})
is plotted in
Fig.\ref{fig:integrands} for typical layers. Consider first a thickness
larger than a few skin depths. One observes a maximum
value of its imaginary part (relevant for the magnetic LDOS) when
the decay constant $\kappa \approx k$ is matched to $1/\delta$. 
This is confirmed by an asymptotic analysis whose results are given
in Table~\ref{t:asymptotics-r}. 
(See Refs.\onlinecite{Henkel99c,Biehs07} for details on the asymptotic
expansion.)  One of the two limiting cases (namely $k_{0} \ll k \ll 1/\delta$)
corresponds precisely to the surface impedance approximation where the
reflection coefficient~(\ref{eq:half-space-r}) is approximated by
\begin{equation}
    r(k ) \approx - 1 + (1+{\rm i}) |k| \delta
    \label{eq:surf-imp-rk}
\end{equation}
Here, the skin depth is much smaller than the lateral period and the 
field barely penetrates into the material.
Fig.\ref{fig:integrands} and Table~\ref{t:asymptotics-r}
show strong deviations in the opposite regime 
$k \gg 1/\delta$ that is relevant at distances $\dist \ll \delta$.

Consider now a layer much thinner than the skin depth. 
>>From Fig.\ref{fig:integrands}, different regimes can be read off
that are separated on the $k$-axis by the scales $\thick / \delta^2 
\ll 1/\thick$, as can be seen in Fig.\ref{fig:integrands}.
It is worth noting that for small $k$, thin layers show even larger 
losses [${\rm Im}\,r(k)$] than thick ones; the maximum is shifted 
towards the smaller value $k \sim \thick / \delta^2$ and has a larger 
amplitude. This
behaviour has been recognized before in magnetic noise 
studies in the kHz range~\cite{Varpula84}. In the infrared range,
it is also well known that the absorption by a metallic layer can
be optimized at a specific thickness.
(See, e.g., Ref.\onlinecite{Bauer92} for incident far-field radiation 
where $|k| \le k_{0}$.) 
Conversely, for a given thickness $\thick$ and dipole distance $\dist$, 
the magnetic noise power shows a maximum as the skin depth is 
changed~\cite{Varpula84,Scheel05a}. This `worst case' occurs when the
characteristic wavevector $1/\dist$ is matched to $\thick / \delta^2$.

\begin{table}[tbp]
    \centering
    \begin{tabular}{|c|c|c|c|}
        \hline
         & $k\ll 1/\delta$ & $1/\delta \ll k$ &\\
        \hline
        $\rule[-2.5ex]{0pt}{6.5ex}
	{\rm Im}\,r_{\rm half\,space}(k)$
	& $k \delta$ & 
	$\displaystyle \frac{ 1 }{ 2 k^2 \delta^2 } $ &\\
        \hline\hline
	& $k \ll 1/\delta$ & $\thick/\delta^2 \ll k \ll 1/\thick $ & $ 1/\thick \ll k$ \\
       \hline
       $\rule[-2.5ex]{0pt}{6.5ex}
       {\rm Im}\,r_{\rm layer}(k)$
       & $\displaystyle{\rm Im} \,
       \frac{-1+(1+{\rm i}) k\delta}{1+{\rm i}k\delta^2/\thick}$
       & $\displaystyle\frac{\thick}{k\delta^2}$
       & $\displaystyle\frac{1}{2k^2\delta^2}$ \\
      \hline
    \end{tabular}
     \caption[]{Asymptotic approximations to the reflection
    coefficients from a half-space and a layer.  We distinguish
    between thin (thickness $\thick \ll \delta$) and thick layers
    ($\thick \ge \delta$, `half space').  The first and second columns
(thin layer) overlap in an intermediate $k$-range (see Fig.\ref{fig:integrands}).
The magnetostatic limit
    $k_{0} \ll k$ is taken throughout.
    These formulas are plotted as dashed lines in
    Fig.\ref{fig:integrands}.
    }
    \label{t:asymptotics-r}
\end{table}


\subsection{Distance dependence}

\begin{figure}[tbp]
    \centering
    \includegraphics*[width=80mm]{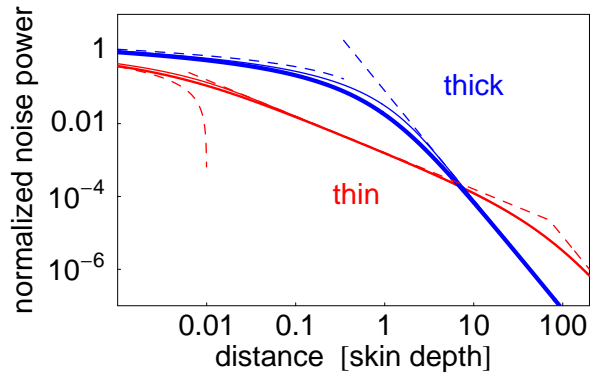}
    \caption[]{Local magnetic noise power ${\cal B}_{ii}( \dist;
    \omega )$ vs.\ distance from 
    metallic layer in double logarithmic scale (2D calculation).
    Top curve [blue]: thick layer; bottom curve [red]: thin 
    layer. The dashed lines give the leading order power 
    laws of Table~\ref{t:power-laws}. The thick curves arise from the
    numerical integration of Eq.(\ref{eq:plane-wave-expansion}),
    the thin curves are an interpolation formula described in the text.
    \\
    The magnetic noise power
    is isotropic above a planar structure in 2D (the perpendicular and 
    parallel field components have the same power). It is scaled to
    $\mu_{0} k_{B}T / (\omega \delta^2)$, and the distance
    is scaled to the skin depth $\delta$.
    Thin [thick] layer:
    $h = 0.01 \,\delta$ [$3 \,\delta$].}
    \label{fig:B-vs-dist}
\end{figure}

The asymptotics in $k$-space translate into power laws for the 
dependence of the magnetic power spectrum $\mathcal{B}_{ii}( \dist ;
\omega )$ on distance $\dist$, as shown in Fig.\ref{fig:B-vs-dist}.
In fact, the integrand in Eq.(\ref{eq:Im-B-yy}) peaks 
around $k \sim 1/(2\dist)$, and the result of the integration is 
determined, to leading order, by the behaviour of $r(k)$ in this 
range. We thus find the power laws summarized in 
Table~\ref{t:power-laws} and visible in
Fig.\ref{fig:B-vs-dist}. We use as
convenient unit in all the plots 
the noise level $\mu_0 k_{\rm B} T / (\omega \delta^2)$. Normalized
to blackbody radiation (in 2D free space), this level is 
$(2/(k_0 \delta))^2 \sim 1.5 \times 10^{12}$ at $1.1\,{\rm MHz}$ for
gold at room temperature, a striking illustration of the Purcell
effect~\cite{Purcell46}. A common trend is that the magnetic noise 
power increases as the metallic medium is approached. As the distance 
$\dist$ is 
getting much smaller than the thickness $\thick$, thin and thick layers
behave the same, as expected. At larger distances, but still
smaller than the skin depth, the noise power is proportional to the 
volume of metallic material, hence to the layer
thickness~\cite{Henkel01a,Folman02}.
This trend is reversed for $\dist > \delta \sqrt{ \delta / 2\thick }$
where thin layers give a larger noise level than thick 
ones~\cite{Varpula84,Scheel05a}. 

\begin{table}[tbp]

    \centering
    \begin{tabular}{|c|c|c|c|}
        \hline
	 & $\dist\ll \delta $  
	 & $\delta \ll \dist$
         &   \\
	\hline
	$\rule[-2.5ex]{0pt}{6.5ex}
	\mathcal{B}_{ii,{\rm half\,space}}(\dist)$
	& $\displaystyle \frac{\log ( \delta / \dist )}{2 \pi}$
	& $\displaystyle\frac{\delta ^{3}}{4 \pi \dist ^{3}}$
	&     \\
      \hline
      \hline
	 & $\dist\ll \thick \ll \delta $
	 & $\thick \ll \dist \ll \delta^2/\thick$
	 & $\delta^2/\thick \ll \dist$ \\
	\hline
	$\rule[-2.5ex]{0pt}{6.5ex}
	\mathcal{B}_{ii,{\rm layer}}(\dist)$
	& $\displaystyle \frac{\log (\thick / \dist)}{2 \pi}$
	& $\displaystyle \frac{\thick}{2 \pi \dist}$
	& $\displaystyle \frac{\delta ^{4}}{4 \pi \thick \dist ^{3}}$   
	\\
      \hline
    \end{tabular}
     \caption[]{Power laws for the magnetic noise spectrum in two
    dimensions above a half space and a thin metallic layer (dashed
    lines in Fig.\ref{fig:B-vs-dist}).  The noise spectrum is given in
    units of $\mu_{0} k_{B}T / (\omega \delta^2)$.}
    \label{t:power-laws}
\end{table}

A reasonably accurate approximation that interpolates between these 
power laws can be found by performing the $k$-integral using the 
asymptotic formulas of Table~\ref{t:asymptotics-r} in their 
respective domains of validity. The result is a sum of 
incomplete gamma functions $\Gamma( n, x, x' )$ ($n = 0, 1, 2$)
whose arguments are, for example, $x \sim \dist \thick /
\delta^2$, $x' \sim \dist / \delta$, or $\dist / \thick$ (thin lines in
Fig.\ref{fig:B-vs-dist} and Appendix~\ref{a:uniform-approximation}).
We have checked that the asymptotics of the gamma function reproduce
the power laws summarized in Table~\ref{t:power-laws}.  There are
regimes where the sub-leading terms give significant corrections, in
particular in the transition regions between the power laws.

Finally, the surface impedance approximation gives a magnetic noise 
that is represented in Fig.\ref{fig:B-vs-dist} by 
the dashed line close to the `thick layer' for
$\dist > \delta$. The agreement with the full calculation in this 
range is expected: the `illuminating field' is getting more and more 
uniform on the scale of the skin depth. At shorter distances, the 
surface impedance approximation severely overestimates the noise level 
because it cannot describe properly the field variations on scales 
smaller than $\delta$. For the thin layer, the conventional surface 
impedance approach gives a wrong result even if $\dist > \delta$
because top and bottom surfaces do not decouple from each other.
This can be repaired using effective
(thickness-dependent) surface impedances, 
see, e.g., Ref.\onlinecite{Tuncer93} and citations therein.

\section{Results: finite size objects}
\label{s:finite-wire}

We now describe numerical calculations that we have performed to
assess the importance of the finite lateral size of the metallic
structure.  This is particularly relevant, for example, in atom chips
where a continuous metallic layer is etched to define wires that can
be addressed with different currents
\cite{Folman02,Reichel02a,Fortagh07}.  It is actually desirable to
minimize the amount of metallic material, leaving just a few wires to
create the fields for atom trapping.  In fact, it has been argued that the
magnetic noise power roughly scales with the metallic volume as long
as the characteristic distances are smaller than the skin depth
\cite{Henkel01a,Folman02}.  For laterally finite structures, this claim
as well as other calculations
have been based so far on approximate methods that fail to
reproduce even the planar layer to within a factor of two or
three~\cite{Henkel01a,Vuletic04,Henkel05c,Dikovsky05}.  The numerical 
results we describe here are a first step towards an accurate 
estimate of magnetic noise power near structures of finite size.

\subsection{Boundary integral equations}

Within the assumption of near field radiation being in equilibrium 
with the metallic object, we compute the noise power from 
the magnetic Green function in Eq.(\ref{eq:FD-theorem}). The magnetic field 
radiated by a point 
source and reflected by the object solves the wave 
equation~(\ref{eq:Maxwell-A-2D}). We reformulate the wave equation
in terms of boundary integral equations.  This has been described elsewhere
\cite{Nieto,Harrington,Rockstuhl03,Rogobete04a}, and we quote only the
basic formulas here.  Our unknowns are the nonzero component $A$ of 
the vector potential and its normal
derivative $F = \partial A / \partial n \equiv
{\bf n} \cdot \nabla A$ on the object surface ${\cal S}$,
where ${\bf n}$ is the outward
unit normal to ${\cal S}$.  Both quantities
are continuous (actually, $F$ is equal to the tangential
magnetic field) and can be found from the system of integral equations
\begin{eqnarray}
    && A( {\bf r} ) =
    2 A_{\rm bulk}( {\bf r} - {\bf r}' ) 
    - \nonumber\\
    &&
    2 \oint\!{\rm d}a( {\bf x} ) \left[
    G_{1}( {\bf r} - {\bf x} ) F( {\bf x} )
    -
    \frac{ \partial G_{1}Ê}{ \partial n }( {\bf r} - {\bf x} ) A( {\bf x} ) 
    \right]
    \label{eq:outside-integral-eqn}
    \\
    && A( {\bf r} ) =
    \nonumber\\
    && 
    2 \oint\!{\rm d}a( {\bf x} ) \left[
    G_{\varepsilon}( {\bf r} - {\bf x} ) F( {\bf x} ) 
    -
    \frac{ \partial G_{\varepsilon} }{ \partial n }
    ( {\bf r} - {\bf x} ) A( {\bf x} )
    \right]
    \label{eq:inside-integral-eqn}
\end{eqnarray}
where $A_{\rm bulk}( {\bf r} - {\bf r}' )$ is given in 
Eq.(\ref{eq:bulk-solution}). Both the observation point ${\bf r}$ 
and the integration points ${\bf x}$ are taken on the 
object boundary ${\cal S}$ here, ${\rm d}a( {\bf x} )$) being
the surface element. We use 
the scalar Green functions [see Eq.(\ref{eq:bulk-solution})]
\begin{equation}
    G_{\varepsilon}( {\bf r} ) = \frac{ {\rm i} }{ 4 }
    H_{0}( k_{0} \sqrt{ \varepsilon } | {\bf r} | )
    .
    \label{eq:def-Green-function}
\end{equation}
If we would take the magnetostatic limit, $G_{1}( {\bf r} ) \to
- (2\pi)^{-1}\log|{\bf r}|$, the Green functions in vacuum and in the medium
would differ (in sub-leading order) by a constant, leading to 
inconsistencies. We avoid this by retaining the finite value of 
$k_{0}$ even for the vacuum Green function. The 
integrals~(\ref{eq:outside-integral-eqn}, \ref{eq:inside-integral-eqn})
are to be understood as principal values to handle the singularities 
of the Green functions as ${\bf x} \to {\bf r}$. We  
discretize them on a finite element decomposition of the object 
boundary, as described in Ref.\onlinecite{Rogobete04a}.
The resulting linear system is solved with standard numerical tools.
Once the fields $A$, $F$ are known on the surface, the reflected field 
at the source position (${\bf r}'$) can be found from
\begin{eqnarray}
    &&
    A_{\rm ref}( {\bf r}' ) = \nonumber
    \\
    &&
    -\oint\!{\rm d}a( {\bf x} ) \left[
    G_{1}( {\bf r}' - {\bf x} ) F( {\bf x} ) 
    -
    \frac{ \partial G_{1} }{ \partial n }( {\bf r}' - {\bf x} ) A( {\bf x} ) 
    \right]
    \hspace*{3em}
    \label{eq:reflected-field}
\end{eqnarray}
Note that $G_{1}$ and $\partial G_{1} / \partial n$ are both
essentially real here (the imaginary parts scale with $k_0$).
The magnetic noise, via ${\rm Im}\,{\bf B}( {\bf r}' )$, is thus
determined by the imaginary parts of $A$ and $F$ on the object
boundary.  This is not surprising since the induced current density is
$\sigma E = {\rm i} \omega \sigma A$.

\subsection{Single wire}

\begin{figure}[tbp]
     \centering
     \includegraphics*[width=85mm]{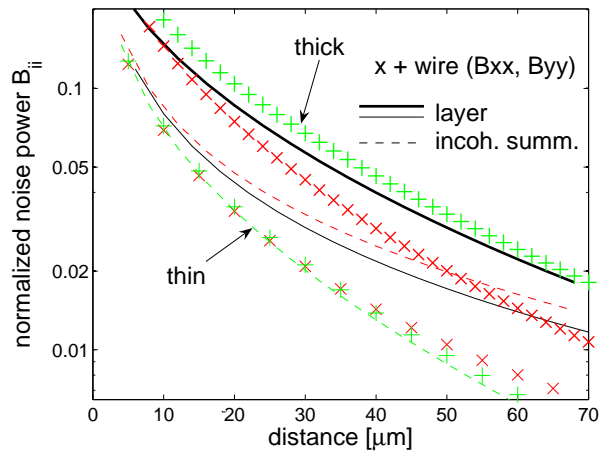}
    \caption[]{Magnetic noise spectra $\mathcal{B}_{ii}( \dist )$
    vs.\ distance $\dist$ above the centre of thin and thick metallic wires.
    Symbols $+$ ($\times$): numerical calculation for
    the $\mathcal{B}_{yy}$ ($\mathcal{B}_{xx}$) component
    (see Fig.\ref{fig:sketch-wire}).
    Solid lines: infinitely wide wire (layer), as computed in
    Sec.\ref{s:layer}.  Dashed lines: incoherent summation
    (thin layer only, upper curve: $\mathcal{B}_{xx}$),
    see Sec.~\ref{s:poor-man}.
    Thin wire: width and thickness $200\times7\,\micron$;
    thick wire: $200\times160\,\micron$.
    The skin depth is $\delta = 70\,\micron$.
	    }
    \label{fig:wide-wires-vs-distance}
\end{figure}

We have solved the integral equations for rectangular wires
of thickness $\thick$ and width $\width$. In a first step, we 
have validated our numerical scheme by comparing flat, wide
wires ($\width \gg \thick$) to the infinite layer results of
Sec.\ref{s:layer}.
Typical plots are shown in Fig.\ref{fig:wide-wires-vs-distance} where 
the magnetic noise power (symbols) is plotted vs.\ the distance $\dist$
above the wire centre. Good agreement with the analytical 
results for an infinitely wide wire (solid lines) is only obtained at 
short distance,
where for geometrical reasons the wire appears wider. At distances
above $20\,\micron$, the deviations start to grow. In all the plots,
we take a skin depth $\delta = 70\,\mu{\rm m}$. 
The slow
convergence in the limit $\width \to \infty$ can be attributed to the 
long-range behaviour of the fields; this is more pronounced in 
two dimensions compared to three. Note in particular 
the strong splitting between the two polarization 
directions for the thick wire that does not occur above an infinitely 
wide wire in 2D (Sec.\ref{s:isotropy}). Interestingly, the $y$-component
shows more noise above a thick wire while this tendency is reversed
above a thin wire.   This polarization
anisotropy could provide a tool to improve the lifetime in a magnetic
trap: one orients the static trapping field parallel to direction of
the strongest noise.  (In fact, trap loss and spin flips are
induced by magnetic fields perpendicular to the static trap field.)
The choice of a trapping field along the weak noise direction is
favorable if one wants to reduce the dephasing rate of the trapped
spin states (generated by fluctuations of the Larmor frequency,
see Ref.\cite{Folman02}).

\begin{figure}[tbp]
    \centering
    \hspace*{-03mm}
    \includegraphics*[width=85mm]{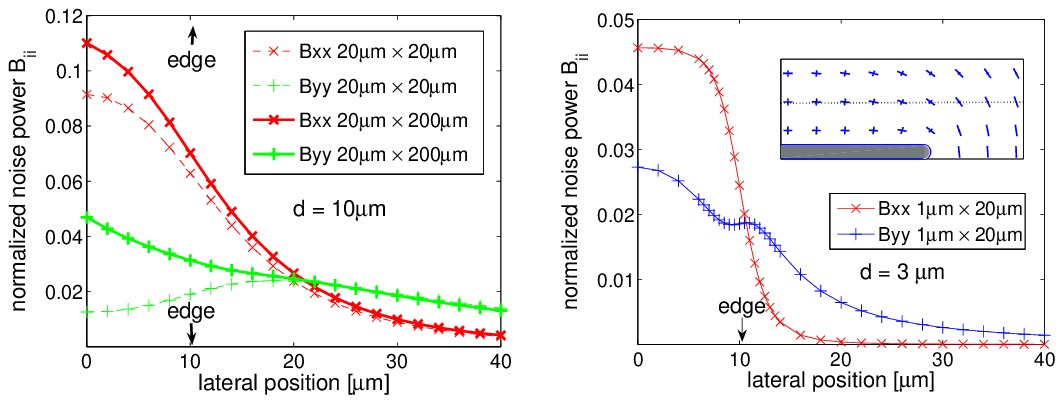}
    \hspace*{-03mm}%
    \caption[]{
    Magnetic noise spectra $\mathcal{B}_{ii}( \dist )$ vs.\
    lateral position, at a fixed distance $\dist$.
    The arrows mark the edges of the wires.  
    Symbols $\times$ [$+$]: spectrum $\mathcal{B}_{xx}$
    [$\mathcal{B}_{yy}$] parallel [perpendicular]
    to the top face of the wire.
    Skin depth: $\delta = 70\,\micron$.
\\
Left panel: thickness and width are
    $20\times200\,\micron$ (wide wire) and $20\times20\,\micron$ (narrow
    wire). Distance $\dist = 10\,\micron$.
Right panel: thickness and width are $1\times20\,\micron$,
    distance $\dist = 3\,\micron$ (see dotted line of inset).
    Inset: illustration of anisotropic noise near the wire edge.  The
    crosses are oriented along the polarization vectors that show
maximum and
    minimum noise, the `arm lengths' being proportional to the rms
    noise. The magnetic field noise is dominantly azimuthal, with
    field lines circling around the wire. The dotted line ($\dist =
3\,\micron$) shows the positions scanned through in the right panel.}
    \label{fig:edge}
\end{figure}

Another finite-size effect is shown in Fig.\ref{fig:edge} where
the position is varied parallel to the top surface of a thin wire. 
Above the centre of wide wires, the noise levels are nearly constant 
(not shown). Beyond the wire edges, one observes a sharp drop in 
${\cal B}_{xx}$, with a characteristic scale fixed by the distance.
The $y$-component shows a broad maximum near the edge that is more
pronounced for narrow wires. 
This is due to a gradually changing direction of maximum noise
that is `azimuthal' with respect to the object, as expected for
magnetic fields generated by currents flowing perpendicular to 
the computational plane (see inset of Fig.\ref{fig:edge}). We find 
the direction of maximum noise by looking for the eigenvectors 
of the $2\times 2$ matrix $\mathcal{B}_{ij}$. 
This matrix can be shown to be symmetric (reciprocity), and 
the eigenvectors that are not orthogonal very near to the wire's
corner, are an artefact of our numerical method that converges 
very slowly at these points.

\begin{figure}[tbp]
    \centering%
    \hspace*{-03mm}%
    \includegraphics*[width=85mm]{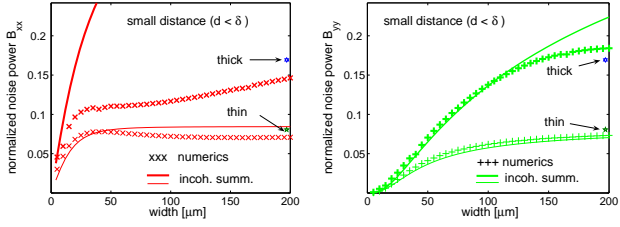}
    \hspace*{-03mm}%
    \caption[]{Magnetic noise spectra vs.\ the width of a rectangular
    wire.  (left) $x$-polarization, parallel to the top face;
(right) $y$-polarization.
Symbols: numerical calculations; solid lines: incoherent summation 
approximation
    (see Sec.\ref{s:poor-man}).
The symbols on the right margin give the values for an
    infinitely wide wire (layer).
The observation point is located
    above the wire centre, at a distance $\dist = 10\,\micron$.  The
    wire thickness is $7\,\micron$ (thin) and $160\,\micron$ (thick).
    Skin depth: $\delta = 70\,\micron$.}
    \label{fig:vs-width-shortdist}
\end{figure}

\begin{figure}[tbp]
     \centering
     \includegraphics*[width=85mm]{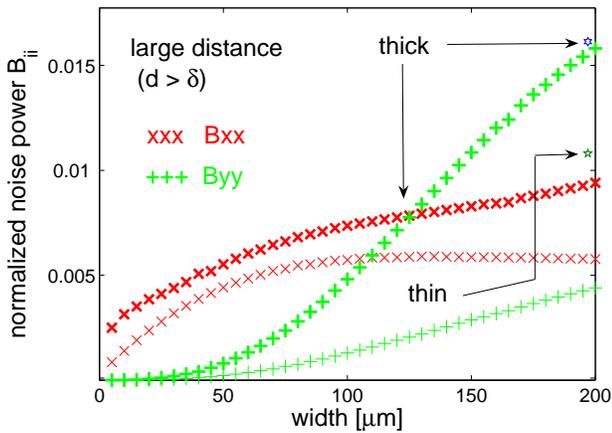}
    \caption[]{Same as Fig.\ref{fig:vs-width-shortdist}, but at
    an observation distance $\dist = 75\,\micron$. Results from the
    incoherent summation are not shown, as they strongly deviate.}
    \label{fig:vs-width-largedist}
\end{figure}

In Figs.\ref{fig:vs-width-shortdist}, \ref{fig:vs-width-largedist},
the thickness of the wire is changed with the observation point 
remaining above the centre. We observe an 
approximately linear increase with the width that saturates slowly. 
We also note that ${\cal B}_{xx}$ (left) levels off faster than
${\cal B}_{yy}$ (right). The difference between 
Fig.\ref{fig:vs-width-shortdist} and Fig.\ref{fig:vs-width-largedist}
is the distance of observation: at short distance
(Fig.\ref{fig:vs-width-shortdist}), the largest widths show a noise 
power fairly close to the planar layer limit (cf.\ the symbols at the
right end). At distances comparable to the skin depth 
(Fig.\ref{fig:vs-width-largedist}), the deviations from the planar 
layer limit (symbols) are still large. Note also that the noise has 
dropped in amplitude and that the increase with width is slower.

\subsection{Incoherent summation}
\label{s:poor-man}

This behaviour can be qualitatively understood using the `incoherent
summation' approximation developed in Ref.\onlinecite{Henkel01a}: the
metallic volume is broken into mutually incoherent point current
elements whose magnetic fields are computed within magnetostatics and
neglecting the presence of the metallic object.
We give the resulting formulas for two dimensions
in Appendix~\ref{a:incoherent-summation}; the integrals are solved 
by special functions for a wire with rectangular cross 
section. The solid lines in Fig.\ref{fig:vs-width-shortdist} 
demonstrate that incoherent summation gives a reliable approximation 
if the skin depth is the largest length scale (not true for the
thick wire). The noise power 
always increases with the metallic volume within this approximation, 
however, and it may also happen
that a wider wire produces
a slightly weaker noise ($x$-polarized curve for a thin wire in
Fig.\ref{fig:vs-width-shortdist}).  This is qualitatively similar to
the trend of Fig.\ref{fig:B-vs-dist} where a thick layer can produce
less noise than a thin one at distances larger than the skin depth.
The polarization anisotropy is also qualitatively reproduced by the 
incoherent summation method, although ${\cal B}_{xx}$ is 
overestimated.  In fact, due to damping on the scale of $\delta$, not
the entire volume of the thick layer contributes to the noise.
The dashed lines in Fig.\ref{fig:wide-wires-vs-distance} and
further calculations show that the quantitative
agreement is systematically better for the field component
perpendicular to the nearest metal surface (here, ${\cal B}_{yy}$).

\subsection{Multiple wires}

\begin{figure}[tbp]
    \centering%
    \hspace*{-03mm}%
    \includegraphics*[width=85mm]{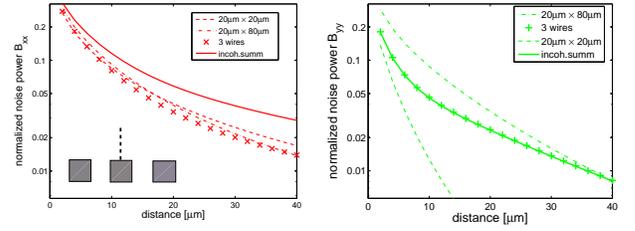}
    \hspace*{-03mm}%
    \caption[]{Noise power generated by three wires, as a function of
    distance (see inserted sketch, with the dashed line illustrating the
    observation points).  The wires have a quadratic cross
    section $20\,\mu{\rm m} \times 20\,\micron$ and are separated by a
    gap of $20\,\mu{\rm m}$.  The noise is measured above the center
    of the central wire.
    Left panel: horizontal polarization, right panel: vertical
    polarization.  Symbols: numerical result; solid line: incoherent
    summation.  For comparison is shown: a single wire of same cross
    section (dashed line) and a wide wire
    $20\,\micron \times 80\,\micron$ with approximately the same
    volume (dash-dotted line).  Skin depth $\delta =
    70\,\micron$.}
    \label{fig:chain-vs-distance}
\end{figure}

This is the generic situation in miniaturized magnetic traps (`atom 
chips') with wires being defined by etchings in a metallic layer. 
We consider three wires of identical cross section and smaller 
than the skin depth. We show in Fig.\ref{fig:chain-vs-distance} the
dependence on the vertical distance, above the central wire. Our results
interpolate smoothly between a single narrow wire ($\dist \ll \width$) and a 
single wide wire ($\dist \agt 2 \width$), as could have been expected.
In fact, the three geometries give nearly the same noise in the azimuthal 
($B_{x}$) polarization. The incoherent summation approximation
overestimates this noise component (similar to 
Fig.\ref{fig:vs-width-shortdist}). We attribute this to correlations 
between the current fluctuations in neighboring wires that are not 
captured by incoherent summation. On the other hand, this 
approximation gives an excellent agreement for the weaker noise component
$B_{y}$.

\begin{figure}[tbp]
    \centering%
    \hspace*{-03mm}%
    \includegraphics*[width=85mm]{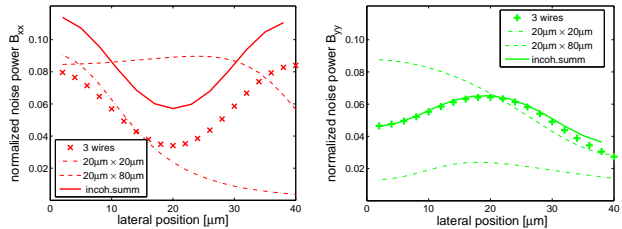}
    \hspace*{-03mm}%
    \caption[]{Same as Fig.\ref{fig:chain-vs-distance}, but at fixed
    distance $\dist = 10\,\micron$, scanning the lateral position.
    $x = 0$ is above the center of the central wire).}
    \label{fig:chain-vs-lateral}
\end{figure}

When we shift the observation point laterally, along the axis connecting
the wire
centres, we get Fig.\ref{fig:chain-vs-lateral}. The stronger $B_{x}$-%
polarization shows maxima of noise above each wire, as expected.  
In the $B_{y}$-polarization, a maximum occurs in the gap between the wires.
This conforms to the general trend of `azimuthal
noise' illustrated in Fig.\ref{fig:edge} (inset).  It is also
interesting that above the central wire ($x = 0$),
three wires generate less noise than only one and also less than 
a single wide wire (approximately a merger of the three). This 
observation goes into the same direction as the experiments reported 
by Nenonen and co-workers~\cite{Nenonen96} where a reduction of
thermal magnetic fields was achieved by cutting a metallic film into
stripes. We attribute this behaviour to negative correlations between 
the currents in neighboring wires brought about by the propagation of
the magnetic field between them. In fact, the noise could only increase
if the wire currents were strictly uncorrelated. 

The performance of the incoherent summation approximation (solid
lines) can be clearly seen, the trends being similar to
Fig.\ref{fig:chain-vs-distance}: good agreement for the
$B_{y}$-polarization, overestimation of the perpendicular case due to
the neglect of correlation effects.

\section{Conclusions}
\label{s:conclusion}

We have described in this paper numerical and analytical results for 
the thermal fields surrounding a two-dimensional metallic object of 
arbitrary cross-section. The role of the skin depth $\delta$ as a 
characteristic length scale has been highlighted. 
At distances smaller than $\delta$, the spectral noise power roughly 
scales with the volume of the metallic material 
(Figs.~\ref{fig:B-vs-dist}, \ref{fig:vs-width-shortdist},
\ref{fig:vs-width-largedist}).  We have reviewed a simple method based
on this idea, the `incoherent summation approximation'.  It 
systematically overestimates the noise power in one of the two 
field polarizations, but otherwise reproduces the main features as
long as the skin depth is the largest scale. 
The strong polarization anisotropy that we have found 
suggests strategies to minimize loss or decoherence due to thermal 
magnetic fields, as observed in recent 
experiments~\cite{Cornell03a,Vuletic04,Zhang05a}: this can be done
by suitably choosing the direction of the static trapping 
fields.  We have also shown that the noise power can be significantly 
non-additive when dealing with multiple objects.  This could be 
relevant for the discrepancy between experiment and theory observed 
in Ref.\onlinecite{Zhang05a}, although our method (restricted to 2D) do 
not permit quantitative predictions of trap lifetimes.

%
%



We now comment on possible extensions of this analysis.
Our framework is also able to provide an approximate
description of superconducting structures. In fact, since we deal with
magnetic field fluctuations at a finite frequency, there is always
some penetration into the superconductor or, equivalently, a finite
resistivity. This can be attributed to a fraction of carriers in a
normally conducting state. Calculations for superconductors of planar 
geometry have been reported in Refs.\onlinecite{Scheel05a,Skagerstam06a},
with applications for miniaturized atom traps in mind. More accurate
descriptions require one to solve the London equations at finite 
frequency inside the superconductor, using for example the two-fluid 
model~\cite{Leung70}.

We recall that we use in this paper a local version of Ohm's law. 
For very pure metallic films, the ballistic 
transport of charge carriers implies a nonlocal 
response~\cite{KliewerFuchs,Feibelman82,Ford84,Koch00}.
This may be particularly relevant for wires defined by doping a 
semiconductor, but would require major changes for the numerical approach 
of Sec.\ref{s:finite-wire}. 

Finally, a brief remark on non-equilibrium settings.  Consider an
isolated metallic object held at a temperature different from its
surroundings (materialized by the vacuum chamber walls, for example).
By applying the generalized Kirchhoff relations (see, e.g., 
Ref.\onlinecite{Rytov3}), the radiation arriving at the observation
point can be split in two parts: one is proportional to the
product of the power a test dipole emits into the far field and the 
temperature of the surroundings; the second part is proportional
to the dipole radiation power absorbed by the object and the metal temperature. 
At the sub-wavelength distances of interest for this paper, one can show 
that the second part is dominant and that the error made in using the
same temperature for metal and surroundings is small. The
equilibrium calculation we have focused on here is then sufficient.

\subsubsection*{Acknowledgements}

This work has been supported by the European Union (project 
ACQP, contract IST-2001-38863) and Universit\"{a}t Potsdam (graduate 
school `Confined Reaction and Interactions in Soft Matter'). We thank 
the quantum optics group in Potsdam for creating a stimulating 
environment.

\appendix

\section{Uniform approximation}
\label{a:uniform-approximation}

The integral over $k$ in Eq.(\ref{eq:Im-B-yy}) can be performed 
analytically if the power-law approximations of Table~\ref{t:asymptotics-r}
are used for the reflection coefficients. We split the integration
range at the crossing points between the power laws and sum the 
contributions. The full expression is cumbersume, and we quote here 
only the most complicated case, the thin layer in the range
$k_{0} \ll k \ll 1/\delta$ (first column of Table~\ref{t:asymptotics-r}).  
The integral can be handled with the formula
\begin{eqnarray}
&&    \int\limits_{k_{1}}^{k_{2}}\!{\rm d}k 
   \, \frac{ (-1+(1+{\rm i})k\delta ) k \,
   {\rm e}^{ - 2 k \dist } }{ 1 + {\rm i} k t }
 \nonumber\\
&& =  (1 - {\rm i}) \frac{ \delta }{ 4 t \dist^2 } 
 \Gamma ( 2, 2 k_{1} \dist, 2 k_{2} \dist ) 
 \label{eq:integral-with-gamma}\\
&& \phantom{=} {}  + \frac {f}{2\dist}\Gamma (1, 2 k_{1} \dist, 2 k_{2} \dist )  
 \nonumber \\
&& \phantom{=} {}  + \frac {{\rm i}\,f}{t} \,Ê{\rm e}^{ -2 {\rm i} d / t} 
    \Gamma (0, 2 k_{1} \dist - 2 {\rm i} \dist / t, 
               2 k_{2} \dist - 2 {\rm i} \dist / t )
	       \nonumber
\end{eqnarray}
where $t = \delta^2 / \thick$ and $f =(1+{\rm i}) \delta^{2} / t
+ {\rm i} / t$. In this case, $k_{1} = k_{0}$ and $k_{2} = (t 
\delta)^{-1/2}$. The incomplete gamma function is defined by
\begin{eqnarray}
    \Gamma( n, b, a ) &=& \Gamma( n, b ) - \Gamma( n, a )
    \label{eq:def-incomplete-gamma}
    \\
    \Gamma( n, a ) &=& \int\limits_{a}^{\infty}\!{\rm d}t \, t^{n-1} 
    {\rm e}^{ - t } 
\end{eqnarray}
Using the asymptotics of the gamma function~\cite{Abramowitz}, we get the 
power laws in Table \ref{t:power-laws}. The logarithmic behaviour 
arises from
\begin{equation}
a \ll 1: \qquad \Gamma (0, a ) \approx -\log a.
\end{equation}

\section{Incoherent summation}
\label{a:incoherent-summation}

We outline here the adaptation of the incoherent summation method of 
Ref.\onlinecite{Henkel01a} to two dimensions. 
The thermal spectrum of the current density 
is given at low frequencies ($\hbar\omega \ll k_{B}T$) by
\begin{equation}
    \langle j^*({\bf x};\omega) j({\bf x}';\omega') \rangle = 
    2\pi\delta( \omega - \omega' )
    2k_{B} T \sigma( {\bf x}; \omega ) \delta( {\bf x} - {\bf x}' )
    \label{eq:}
\end{equation}
This spectrum is already integrated over a unit length in the
$z$-direction (parallel to the current) along which the current
density is assumed to be uniform (two-dimensional geometry).  In this
formulation, $\sigma$ is (the real part of) the 3D conductivity that
we assume local, as reflected by the spatial $\delta$-correlation.  We
only take into account currents parallel to the $z$-direction.  Each
current element generates a magnetic field in the $xy$-plane that we
compute in the magnetostatic approximation and ignoring the presence
of the embedding metal.  The latter point is the key approximation
made.  This gives a magnetic noise spectrum (integrated over a unit
length along $z$) of the order of
\begin{equation}
    S_{B} = 
    \frac{ \mu_{0} k_{B} T }{ 
    \omega \delta^2 }
    ,
    \label{eq:def-spectrum}
\end{equation}
with cross correlations given by
\begin{eqnarray}
    \mathcal{B}_{ij}( {\bf x}; \omega ) = 
    \frac{ S_{B} }{ \pi^2 } 
    \left( \delta_{ij} ({\rm tr}\,Y) - Y_{ij} \right)
    \label{eq:magnetic-power}
    \\
    Y_{ij}( {\bf x} ) = 
    \int\limits_{V}\!{\rm d}x_{1}'{\rm d}x_{2}'
    \frac{
    (x_{i} - x_{i}' ) (x_{j} - x_{j}' )
    }{ | {\bf x} - {\bf x}' |^4 }
    \label{eq:def-Yij}
\end{eqnarray}
where $V$ is the volume occupied by the metal. The `geometrical tensor' 
$Y_{ij}$ is dimensionless (a specific 2D property) and depends only on 
the ratio of observation distance and object size. It does not involve 
the skin depth, of course.

For a microstructure with rectangular cross section, an observer 
located above the center of the structure sees a noise power
\begin{eqnarray}
    \frac{ \mathcal{B}_{xx}( \dist; \omega ) }{
    S_{B} }
    &=& \frac {1}{2\pi^2} 
    \left[ \left[ \arctan(\frac {x'}{y-y'})\right]^{\frac{w}{2}}
         _{x'=-\frac{w}{2}} \right]^{0}_{y'=-h} 
    \nonumber
    \\
    && + \frac {1}{2\pi^2}\left[ \left[ 
    {\rm Im} \,{\rm Li}_{2}(\frac {{\rm i} x'}{y-y'} ) \right]
    ^{\frac{w}{2}}_{x'=-\frac {w}{2}} \right] ^{0}_{y'=-h}
\\
\frac{ \mathcal{B}_{yy}( \dist; \omega ) }{
    S_{B} }
    &=& \frac {1}{2\pi^2} 
    \left[ \left[ \arctan(\frac {y-y'}{x'})
    \right]^{\frac{w}{2}}_{x'=-\frac{w}{2}}     
    \right]^{0}_{y'=-h} 
    \nonumber
    \\
    && + \frac {1}{2\pi^2}\left[ \left[ 
    {\rm Im} \,{\rm Li}_{2}(\frac {{\rm i} x'}{y-y'} ) 
    \right]^{\frac{w}{2}}_{x'=-\frac {w}{2}} 
    \right]^{0}_{y'=-h}
    \nonumber
\end{eqnarray}
where ${\rm Li}_{n}( \cdot )$ is the polylogarithm and we 
have used the notation
\begin{equation}
\left[ \left[ f(u,v) \right]^{b}_{u=a} \right]^{d}_{v=c} \equiv 
f(a,c)-f(a,d)-f(b,c)+f(b,d)
\end{equation}

%
%


\end{document}